\documentclass{aa}
\usepackage{times}
\usepackage{graphics}
\usepackage{xspace}
\usepackage{epsfig}
\usepackage{rotating}
\usepackage{dcolumn}

\makeatletter
\@ifundefined{chapter}{\def\thebibliography#1{
\section*{References}                          
  \list
  {\relax}{\setlength{\labelsep}{0em}
        \setlength{\itemindent}{-\bibhang}
        \setlength{\itemsep}{\parskip}          
        \setlength{\parsep}{0pt}
        \setlength{\leftmargin}{\bibhang}}
    \def\newblock{\hskip .11em plus .33em minus .07em}
    \sloppy\clubpenalty4000\widowpenalty4000
    \sfcode`\.=1000\relax}}%
{\def\thebibliography#1{
  \list
  {\relax}{\setlength{\labelsep}{0em}
        \setlength{\itemindent}{-\bibhang}
        \setlength{\itemsep}{\parskip}          
        \setlength{\parsep}{0pt}
        \setlength{\leftmargin}{\bibhang}}
    \def\newblock{\hskip .11em plus .33em minus .07em}
    \sloppy\clubpenalty4000\widowpenalty4000
    \sfcode`\.=1000\relax}}

\newlength{\bibhang}
\let\@internalcite\cite
\def\cite{\@ifstar{\citey}{\citefull}}
\def\citefull{\def\astroncite##1##2{##1\ ##2}\@internalcite}
\def\citey{\def\astroncite##1##2{##1\ (##2)}\@internalcite}
\def\citeyear{\def\astroncite##1##2{##2}\@internalcite}
\def\citename{\def\astroncite##1##2{##1}\@internalcite}
\def\@citex[#1]#2{\if@filesw\immediate\write\@auxout{\string\citation{#2}}\fi
  \def\@citea{}\@cite{\@for\@citeb:=#2\do
    {\@citea\def\@citea{; }\@ifundefined       
       {b@\@citeb}{{\bf ??}\@warning              
       {Citation `\@citeb' on page \thepage \space undefined}}%
{\csname b@\@citeb\endcsname}}}{#1}}

\def\@cite#1#2{#1\if@tempswa #2\fi}       
\def\@biblabel#1{}

\def\astroncite#1#2{#1\ #2}
\makeatother

\newcommand{\newrule}{\rule[-0.05cm]{0.cm}{0.4cm}}

\begin{document}

\thesaurus{06(%
13.25.5; 
08.06.2)}

\title{X-ray emission from young stars in the Tucanae association}

\author{B. Stelzer \and R. Neuh\"auser} 

\institute{Max-Planck-Institut f\"ur extraterrestrische Physik,
  Giessenbachstr.~1,
  D-85740 Garching,
  Germany} 

\offprints{B. Stelzer}
\mail{B. Stelzer, stelzer@xray.mpe.mpg.de}
\titlerunning{X-rays from the Tucanae association}

\date{Received $<$17 April 2000$>$ / Accepted $<$14 July 2000$>$ } 
\maketitle

\begin{abstract}

We report on X-ray emission from members of the recently discovered
Tucanae association, a group of stars with youth signatures and similar
space motion. The Tucanae association is the nearest known 
region of recent star formation ($\sim 45\,{\rm pc}$) 
far from molecular clouds (\cite{Zuckerman00.1}).
We have made use of the {\em ROSAT} Data Archive and searched for
X-rays from Tucanae stars in both {\em ROSAT} All-Sky Survey (RASS) and
pointed observations. While the RASS provides complete coverage of the
sky, only three potential Tucanae members have been observed
during PSPC pointings. All three stars have been detected. For the RASS the
percentage of detections is 59\%.

The comparison of the X-ray luminosity function of Tucanae 
to that of other star forming regions may provide clues to the uncertain 
age of the association. We find that the distribution of X-ray 
luminosities is very similar to the ones derived for the 
TW\,Hya association, the Taurus-Auriga T Tauri Stars, and the IC\,2602 cluster,
but significantly brighter than the luminosity distribution of the
Pleiades. We conclude that the stars in Tucanae are most likely young, 
on the order of $10-30\,{\rm Myr}$.

Strong variability of most stars emerges from the X-ray lightcurves
where several flares and irregular variations are observed.


\end{abstract}

\newlength{\mylength}

\settowidth{\mylength}{Melotte22-2147}

\section{Introduction}\label{sect:intro}

The discovery of the TW\,Hydrae association as a nearby 
group of co-moving T Tauri Stars (TTS) 
far from molecular clouds has motivated the
search for additional associations of pre-main sequence stars (PMS) 
close to Earth.
In the course of a study of the {\em Hipparcos} catalog around 
IRAS 60$\mu{\rm m}$ sources \citey{Zuckerman00.1} recently announced
the identification of a co-moving group of stars with general youth
indicators.

Youth can be deduced from the presence of the $6708\,{\rm \AA}$ 
Li absorption line, H$\alpha$
profile, rotation rate, IR excess indicative of circumstellar material, 
and X-ray emission. While any of these properties certainly are not
sufficient to establish the youth of a star, the common presence of several
of these characteristics is suggestive for PMS nature. 
However, no accurate age determination is possible by these means.
In addition, for all but one probable Tucanae members, {\em Hipparcos} 
parallaxes are available (see \cite{Zuckerman00.1} and our tables~2~and~3),
so that the stars can be placed accurately into the H-R diagram. They all
fall near and/or above the zero-age main sequence, hence are really young.

Based on proper motion, distance, and the above mentioned youth indicators
\citey{Zuckerman00.1} have divided their sample of potential Tucanae
members in two groups: probable members with common distance, space motion,
and general signs of youth, and improbable members. This
latter group consists of stars which either could be members
without signs of youth, or stars which by chance have distances and proper
motion similar to the association. \citey{Zuckerman00.1} have identified
11 of the probable members with X-ray sources from the {\em ROSAT} All-Sky
Survey Bright Source
Catalog (RASS BSC; \cite{Voges99.1}), but only one of the non-members.

In this paper we present a more detailed study of the X-ray emission
of Tucanae candidates based on data obtained from the {\em ROSAT}
Public Data Archives. In particular we use X-ray luminosity distribution
functions (XLDFs) to obtain further clues to the age of this association.
Comparative studies of young stellar clusters and star-forming regions 
based on observations by the {\em Einstein} IPC (see
e.g. \cite{Feigelson89.1}, \cite{Damiani95.1})
have shown that the X-ray luminosity 
decays with stellar age. This is manifest in a decrease of the median of 
$L_{\rm x}$ and a corresponding 
shift of the XLDFs with respect to each other. A
comparison of the XLDF for the Tucanae stars 
with that for well studied star forming regions should therefore allow to put
constraints on the age of this association. 

In Sect.~\ref{sect:obs} we describe the observations and analysis of the
raw data. We also provide tables which summarize the X-ray properties
of detected and undetected potential Tucanae members. The XLDFs of
Tucanae candidates and comparison samples are discussed in 
Sect.~\ref{sect:xldfs}. In Sect.~\ref{sect:variability} we present the
X-ray lightcurves of all detected stars and discuss their variability.
Our results are summarized in Sect.~\ref{sect:conclusions}.

\section{Observations}\label{sect:obs}

This study is based on the list of potential members of the Tucanae
association provided by \citey{Zuckerman00.1}. We analyze the {\em ROSAT}
observations of all stars
from their Table~1, i.e. both `probable' members (Table~1a) 
and stars termed `improbable' members (Table~1b).

The {\em ROSAT} satellite is equipped with two X-ray detectors in the focus
of its telescope, the Position Sensitive Proportional Counter (PSPC) and
the High Resolution Imager (HRI). See \citey{Truemper83.1},
\citey{Pfeffermann88.1} and \citey{David99.1} for a description of the 
instrumentation for the {\em ROSAT} mission. 
The raw data of all observations can be retrieved from 
public archives. Our cross-correlation of Table~1 from \citey{Zuckerman00.1}
with the archive showed that none of the stars was in the field
of any HRI exposure. However, three stars have been observed in pointed
observations of the PSPC. All stars in Tucanae have been observed during
the {\em ROSAT} All Sky Survey (RASS).

In this section we describe our analysis of the PSPC data (in both pointed
and survey mode) using the Extended Scientific Analysis System 
(EXSAS; \cite{Zimmermann95.1}). The results of source detection are
presented and the properties of detected and undetected stars are
summarized.

\subsection{PSPC pointed observations}\label{subsect:pspc}

Three stars from the Tucanae membership lists are in the field of view of 
a pointed PSPC observation (HIP\,92680, HIP\,100751, and HIP\,103438). 
The {\em ROSAT} observation request numbers (ROR) 
are 201597p, 200099p, and 200404p, respectively.
After extracting the raw data from the
archive we have performed source detection on these exposures 
using a combined local and map detection algorithm 
based on a maximum likelihood technique (\cite{Cruddace88.1}). 
For the cross-correlation of detected
X-ray sources with the membership list introduced above 
we allow a maximum distance between
X-ray and optical position of $40^{\prime\prime}$ (shown by
\cite{Neuhaeuser95.1} to minimize
the contamination by background sources).
All stars are clearly identified with an X-ray source at an offset 
less than $30^{\prime\prime}$ from the optical position.

The number of counts in the background map at the source location
is scaled to the photon extraction area and subtracted from the total 
number of counts.
Count rates are computed using the information in the exposure maps, and
transformed into luminosities applying the individual distances of the stars 
(derived from the {\em Hipparcos} parallax) and an energy-conversion-factor
(ECF) determined from the hardness ratio. 

Hardness ratios computed from the standard energy bands of the PSPC 
provide spectral information in case of low signal-to-noise.
The PSPC hardness ratios are defined as follows 
\begin{equation}
HR1 = \frac{H - S}{H + S}
\label{eq:hr1}
\end{equation}
\begin{equation}
HR2 = \frac{H2 - H1}{H2 + H1}
\label{eq:hr2}
\end{equation}
where $H$, $S$, $H2$, and $H1$ substitute the number of counts in the
respective energy band: $S = 0.1-0.4\,{\rm keV}$, $H1 = 0.5-0.9\,{\rm
keV}$, $H2 = 0.9-2.0\,{\rm keV}$, and $H = H1 + H2$. 

As discussed by \citey{Fleming95.1} for late-type stars 
the $ECF$ can be computed from $HR1$ and is given by
\begin{equation}
ECF = (8.31 + 5.30 HR1)\,10^{-12}\,{\rm erg\,cm^{-2}\,cts^{-1}}
\label{eq:ecf}
\end{equation}
For early-type stars (HIP\,100751 has spectral type B) this conversion may
not be appropriate. In this case we have assumed 
thermal emission at $kT = 0.5\,{\rm keV}$ (see \cite{Berghoefer96.1}) 
and negligible absorption and
computed the $ECF$ following the Technical Appendix F to the {\em ROSAT}
Call for Proposals. Alternatively, count rates have been converted
to fluxes using standard EXSAS routines. The luminosities derived by the 
two methods were found to be in good agreement, showing that the Fleming
relation holds for the Tucanae stars.

The pointed PSPC observations of stars from our membership list 
are summarized in 
Table~\ref{tab:allpspc}. Next to the {\em Hipparcos} number, X-ray position,
and offset between optical and X-ray position $\Delta$ (column~4) we give
the maximum likelihood of existence (column~5), the exposure time
(column~6), and the broad band count rate (column~7). 
To derive the X-ray luminosity
(shown in column~8) we have divided the count rate by the number of
components in the system (HIP\,103438 is a binary, the other two stars are
singles; see also Tables~\ref{tab:all_sol}~and~\ref{tab:all_upp} for the
multiplicity of the Tucanae stars). 
This implies that all members contribute the same
amount of X-ray emission and was 
shown by \citey{Koenig00.1} to be acceptable in
almost all cases according to {\em ROSAT} HRI observations of resolved
young binary stars in Taurus. 
Finally, columns~9~and~10 contain the hardness ratios. 

\begin{table*}
\begin{center}
\caption{X-ray properties of stars in the Tucanae region as observed during
pointed PSPC observations. Only three stars have been observed including
HIP\,103438, a star from the list of improbable members. 
The {\em ROSAT} request numbers of the
respective observations are 201597p (HIP\,92680), 200099p (HIP\,100751),
and 200404p (HIP\,103438). Note, that HIP\,103438 is a binary, and we have
divided the observed count rate by two to obtain $L_{\rm x}$. See text for
a description of the entries in individual columns.}
\label{tab:allpspc}
\begin{tabular}{lrrrrrrrrr}\hline
HIP & \multicolumn{1}{c}{$\alpha_{\rm 2000}$} &
              \multicolumn{1}{c}{$\delta_{\rm 2000}$} &
              \multicolumn{1}{c}{$\Delta$} & 
              \multicolumn{1}{c}{Exi ML} &
              \multicolumn{1}{c}{Expo} & \multicolumn{1}{c}{Rate} &
              \multicolumn{1}{c}{$L_{\rm x}$} & \multicolumn{1}{c}{$HR1$} & \multicolumn{1}{c}{$HR2$} \\ 
              & \multicolumn{1}{c}{[hh:mm:ss]} &
              \multicolumn{1}{c}{[dd:mm:ss]} &
              \multicolumn{1}{c}{[$^{\prime\prime}$]} &  &
              \multicolumn{1}{c}{[s]} & \multicolumn{1}{c}{[cts/s]} &
              \multicolumn{1}{c}{[$10^{29}\,{\rm erg/s}$]} & & \\ \hline
 92680        &        \newrule       18 53 05.2      &       -50 10 48
 &          7.1  &          28078        &          22398        &       $
      1.142      \pm          0.007      $       &       $          28.8
 \pm         0.8         $       &       $        0.05   \pm      0.01   $
 &       $        0.00   \pm      0.01   $        \\ 
 100751       &        \newrule       20 25 38.4      &       -56 44 05
 &          3.8  &            101        &           1098        &       $
      0.034      \pm          0.006      $       &       $           1.1
 \pm         0.2         $       &       $       -0.23   \pm      0.16   $
 &       $       -0.07   \pm      0.26   $        \\ 
 103438       &        \newrule       20 57 25.1      &       -59 04 17
 &         26.8  &             39        &           3256        &       $
      0.037      \pm          0.005      $       &       $           0.3
 \pm         0.2         $       &       $       -0.66   \pm      0.13   $
 &       $       -0.36   \pm      0.46   $        \\ \hline
\end{tabular}
\end{center}
\end{table*}

\subsection{RASS observations}\label{subsect:rass}

During the first months of its operation {\em ROSAT} performed an 
All-Sky Survey.
In the course of this program the whole sky was scanned by the $2^\circ$
field of view of the PSPC. The FOV was shifted by $4^\prime$ per scan,
such that the total number of scans of a specific location in the
sky is $\sim$ 25. The exposure times depend on the ecliptic latitude
$\beta$ and scale with $1/\cos{\beta}$.
For the Tucanae stars they range from $\sim 50-500$ s.

Owing to the short exposures ($\sim$ 30\,s per scan) the sensitivity
of the RASS is limited to $\sim 2\,10^{28}\,{\rm erg/s}$ 
at a distance of 45\,pc.
Its main advantage is therefore the spatial completeness. In contrast to the
pointed observations, where only three Tucanae candidates have been observed, 
X-ray data for all potential Tucanae stars are available from the RASS.

We have retrieved the RASS raw data from the 
Public Archive. For the source detection we proceed in a similar way as
described in Sect.~\ref{subsect:pspc} for pointed PSPC observations. 
The maximum offset allowed between
optical and X-ray position is again $40^{\prime\prime}$.
As a consequence of the scanning mode a given source has a different 
off-axis angle in each RASS scan. 
The compilation of count rates and luminosities follows the description 
outlined in Sect.~\ref{subsect:pspc}.
Since for non-detections no information about the spectral hardness is
available, we use the mean of the $ECF$ of detected Tucanae members for the
conversion of upper limit counts to upper limit $L_{\rm x}$.

The detection fraction is much larger among the likely members of Tucanae
(13/22) than in the group of improbable members (1/15). The X-ray
properties of all detected and undetected stars are given in 
Tables~\ref{tab:all_sol}~and~\ref{tab:all_upp}. The meaning of columns~1~to~7
in Table~\ref{tab:all_sol} (detected sources) is the same as in 
Table~\ref{tab:allpspc}. 
We provide the distance derived
from the {\em Hipparcos} parallax in column~8. The multiplicity of the
stellar system is given in column~9, and the spectral type in
column~10. The multiplicity is used to derive the X-ray 
luminosity (column~11) as described in 
Sect.~\ref{subsect:pspc}. The last three columns show the ratio of X-ray
to bolometric luminosity, and the PSPC hardness ratios.
$L_{\rm bol}$ was derived from the $V$ magnitude and spectral type
(as given by \cite{Zuckerman00.1}) using the bolometric correction of
\citey{Schmidt-Kaler82.1} and assuming negligible absorption. This is
justified because of the small distance and lack of intervening gas in
the line of sight. The low ($\simeq 0.0$) hardness ratios
of the Tucanae stars also show that absorption is small if not negligible.
Note, that no individual $V$ magnitudes are available for the components in
close multiple
systems. Therefore, we have used the combined $L_{\rm x}$ as well as the
combined $L_{\rm bol}$ (i.e. without
scaling to the number of components) in order to compute 
$L_{\rm x}/L_{\rm bol}$. 

For undetected sources (Table~\ref{tab:all_upp}) we
list the {\em Hipparcos} number, 
the optical position (columns~2~and~3), exposure time (column~4), 
upper limit to the broad band count rate (column~5), 
the distance to the star (column~6), multiplicity (column~7), 
spectral type (column~8), 
and upper limits to the X-ray luminosity (column~9) and to the 
$L_{\rm x}/L_{\rm bol}$-ratio (column~10). 
As for detected sources $L_{\rm x}$ has been computed by 
dividing the observed count rate by the multiplicity.
\begin{sidewaystable*}
\begin{center}
\caption{RASS X-ray data for candidate members of the Tucanae
association, i.e. stars from Table~1 in \protect\citey{Zuckerman00.1}. We
give the designations from the {\em Hipparcos} catalogue. One star has no
{\em Hipparcos} parallax and we list its PPM number. The X-ray
position (column~2~and~3), distance to optical position $\Delta$ (column~4),
maximum likelihood of existence (column~5), exposure time (column~6), and
broad band count rate (column~7) are given. The distance (column~8) is
derived from the {\em Hipparcos} parallax. (For PPM366328 no parallax has
been measured and the guess by \protect\cite{Zuckerman00.1} has been used for
the distance.) The spectral types listed in column~10 are from
\protect\citey{Zuckerman00.1}. $L_{\rm x}$ in column~11 
is computed under the assumption that all components in multiples emit the 
same amount of X-rays, i.e. we have divided the counts by the multiplicity
(given in column~9). Since for close multiples only one value of the combined
$V$ magnitude is available, the X-ray luminosity used to compute 
$\lg{(L_{\rm x}/L_{\rm bol})}$ (column~12) is the observed value, i.e. 
the combined luminosity without taking account of the number of components
in the system. Columns~13~and~14 contain the PSPC hardness ratios.}
\label{tab:all_sol}
\begin{tabular}{lrrrrrrrclrrrr}\hline
HIP & \multicolumn{2}{c}{X-ray position} &
              \multicolumn{1}{c}{$\Delta$} & \multicolumn{1}{c}{Exi ML} &
              \multicolumn{1}{c}{Expo} & \multicolumn{1}{c}{Rate} &
              \multicolumn{1}{c}{distance} & \multicolumn{1}{c}{Multi} &
              \multicolumn{1}{c}{Sp.Type} & \multicolumn{1}{c}{$L_{\rm x}$} & \multicolumn{1}{c}{$\lg{}$} & \multicolumn{1}{c}{$HR1$} & \multicolumn{1}{c}{$HR2$
} \\ 
              & \multicolumn{1}{c}{[R.A.: h\,m\,s]} &
              \multicolumn{1}{c}{[Dec.: d\,m\,s]} &
              \multicolumn{1}{c}{[$^{\prime\prime}$]} & &
              \multicolumn{1}{c}{[s]} & \multicolumn{1}{c}{[cts/s]} &
              \multicolumn{1}{c}{[pc]} & & &
              \multicolumn{1}{c}{[$10^{29}\,{\rm erg/s}$]} & $L_{\rm x}/L_{\rm bol}$ & & \\ \hline
\multicolumn{14}{c}{\em Probable Tuc members} \newrule \\ \hline
 1481         &        \newrule       00 18 26.9      &       -63 28 39 &          7.4  &           111.3       &          195.1        &       $      0.289      \pm          0.042      $       &          40.9         & 1       &       F8      &       $           4.2         \pm         0.8 $       &          -4.02        &       $       -0.19   \pm      0.14   $ &       $       -0.10   \pm      0.23   $        \\ 
 1910         &        \newrule       00 24 07.8      &       -62 10 58 &          2.4  &            19.9       &           97.6        &       $      0.155      \pm          0.045      $       &          46.3         & 1       &       M0      &       $           2.8         \pm         1.0 $       &          -2.30        &       $       -0.26   \pm      0.27   $ &       $       -0.24   \pm      0.48   $        \\ 
 2729         &        \newrule       00 34 53.2      &       -61 55 09 &          9.7  &           156.8       &          142.5        &       $      0.523      \pm          0.064      $       &          45.9         & 1       &       K4      &       $           9.7         \pm         1.8 $       &          -2.76        &       $       -0.19   \pm      0.12   $ &       $       -0.12   \pm      0.20   $        \\ 
 92680        &        \newrule       18 53 06.0      &       -50 10 42 &          7.1  &           390.6       &          143.3        &       $      0.986      \pm          0.086      $       &          49.6         & 1       &       K0      &       $          23.8         \pm         2.1 $       &          -3.25        &       $       -0.02   \pm      0.09   $ &       $        0.12   \pm      0.12   $        \\ 
 93815        &        \newrule       19 06 20.0      &       -52 20 24 &         10.7  &          1052.1       &          134.9        &       $      2.055      \pm          0.126      $       &          52.4         & 2       &       F7      &       $          31.8         \pm         4.2 $       &          -3.98        &       $        0.21   \pm      0.06   $ &       $        0.31   \pm      0.08   $        \\ 
 99803        &        \newrule       20 14 54.9      &       -56 58 29 &          3.8  &            26.6       &          176.7        &       $      0.127      \pm          0.033      $       &          65.2         & 1       &       F6.5    &       $           6.1         \pm         1.7 $       &          -4.37        &       $        0.21   \pm      0.26   $ &       $       -0.63   \pm      0.28   $        \\ 
 105388       &        \newrule       21 20 50.7      &       -53 01 56 &              9.9  &           563.3       &          380.7   &   $ 0.645      \pm          0.043      $       &          45.8         & 1       &       G5      &       $          12.4         \pm         1.4 $       &          -3.19        &       $       -0.13   \pm      0.07   $ &       $       -0.12   \pm      0.10   $        \\ 
 105404       &        \newrule       21 21 00.1      &       -52 28 33 &         18.4  &           717.7       &          372.4        &       $      0.795      \pm          0.048      $       &          46.0         & 2       &       K0      &       $           8.1         \pm         0.6 $       &          -2.91        &       $       -0.04   \pm      0.06   $ &       $        0.10   \pm      0.09   $        \\ 
 107345       &        \newrule       21 44 30.0      &       -60 58 29 &          9.5  &            70.4       &          329.3        &       $      0.152      \pm          0.024      $       &          42.2         & 1       &       M1      &       $           2.4         \pm         0.5 $       &          -2.03        &       $       -0.20   \pm      0.16   $ &       $       -0.05   \pm      0.25   $        \\ 
 107947       &        \newrule       21 52 10.2      &       -62 03 08 &          3.2  &           396.7       &          389.2        &       $      0.448      \pm          0.036      $       &          45.0         & 1       &       F6      &       $           8.8         \pm         0.8 $       &          -3.91        &       $       -0.05   \pm      0.08   $ &       $        0.14   \pm      0.12   $        \\ 
 108195       &        \newrule       21 55 12.8      &       -61 53 20 &         13.6  &           130.2       &          445.3        &       $      0.190      \pm          0.023      $       &          46.5         & 2       &       F3      &       $           2.0         \pm         0.2 $       &          -4.82        &       $       -0.06   \pm      0.12   $ &       $       -0.10   \pm      0.17   $        \\ 
 PPM\,366328       &        \newrule       23 15 00.6      &       -63 34 32 &          8.1  &           129.1       &          501.3        &       $      0.159      \pm          0.020      $       &          50.0         & 1       &       K0      &       $           3.7         \pm         0.5 $       &          -3.26        &       $       -0.09   \pm      0.12   $ &       $       -0.20   \pm      0.18   $        \\ 
 116748       &        \newrule       23 39 40.1      &       -69 11 47 &              4.9  &          1187.5       &          475.  & $    0.909      \pm          0.045      $       &          46.2         & 2       &       G5+8    &       $           8.7         \pm         1.0 $       &          -3.20        &       $       -0.15   \pm      0.05   $ &       $        0.11   \pm      0.08   $        \\ \hline
\multicolumn{14}{c}{\em Improbable Tuc members} \newrule \\ \hline
 103438       &        \newrule       20 57 20.6      &       -59 04 16 &         22.1  &            21.8       &          291.2        &       $      0.067      \pm          0.018      $       &          50.9         & 2       &       G2    &       $           0.5         \pm         0.4 $       &          -4.65        &       $       -0.69   \pm      0.22   $ &       $       -0.30   \pm      0.68   $        \\ \hline
\end{tabular}
\end{center}
\end{sidewaystable*}

\begin{table*}
\begin{center}
\caption{RASS X-ray data for undetected candidate members of the Tucanae
association. Designations in column~1 are the {\em Hipparcos}
numbers. Upper limits have been measured at the optical 
position of the stars (see columns~2~and~3). Column~4 contains the exposure
time, and column~5 the upper limit to the broad band count rate. The
distance derived from the {\em Hipparcos} parallax is given in
column~6. The spectral types listed in column~8 are from
\protect\citey{Zuckerman00.1}. The multiplicity given in column~7 was used to compute the X-ray
luminosity (column~9) for the individual components in multiples as
described in the text. Note, that HIP\,2484 and HIP\,2487 build a 
triple but are 
listed separately because their optical position is different. For the
compilation of the $L_{\rm x}/L_{\rm bol}$ ratio 
given in column~10 we have used the observed count rates, i.e. the combined
luminosity of all components in case of multiples since only combined
$V$ magnitudes are available.}
\label{tab:all_upp}
\begin{tabular}{lrrrrrclrr}\hline
HIP & \multicolumn{2}{c}{Position} & Expo & \multicolumn{1}{c}{Rate}
              & Distance         & \multicolumn{1}{c}{Multi} & Sp.Type & \multicolumn{1}{c}{$\lg{L_{\rm x}}$} & $\lg{(L_{\rm x}/L_{\rm bol})}$ \\ 
              & [R.A.: h\,m\,s]        & [Dec.: d\,m\,s]   & \multicolumn{1}{c}{[s]}  &       [$10^{-3}$\,cts/s] & [pc]     &    &    & [${\rm erg/s}$] & \\ \hline
\multicolumn{10}{c}{\em Probable Tuc members} \newrule \\ \hline
 1993         &        \newrule       00 25 14.6      &       -61 30 48
 &           50.5        &       $        <          46.0        $       &
    37.4         &       1       &       K7      &       $        <        28.79
         $       &       $        <       -2.91  $        \\ 
 2484         &        \newrule       00 31 32.6      &       -62 57 29
 &          105.9        &       $        <          37.2        $       &
    42.8         &       1       &       B9      &       $        <        28.81
         $       &       $        <       -5.98  $        \\ 
 2487         &        \newrule       00 31 33.4      &       -62 57 56
 &          105.9        &       $        <          39.4        $       &
    52.7         &       2       &       A2+7    &       $        <        28.72
         $       &       $        <       -6.01  $        \\ 
 2578         &        \newrule       00 32 43.8      &       -63 01 53
 &          100.4        &       $        <           4.8        $       &
    46.4         &       1       &       A0      &       $        <        28.00
         $       &       $        <       -6.66  $        \\ 
 95261        &        \newrule       19 22 51.2      &       -54 25 25
 &           61.7        &       $        <          42.8        $       &
    47.6         &       1       &       A0      &       $        <        28.97
         $       &       $        <       -5.73  $        \\ 
 95270        &        \newrule       19 22 58.9      &       -54 32 16
 &           61.7        &       $        <         102.7        $       &
    50.5         &       1       &       F5      &       $        <        29.40
         $       &       $        <       -4.63  $        \\ 
 100751       &        \newrule       20 25 38.8      &       -56 44 06
 &          265.5        &       $        <          13.6        $       &
    56.8         &       1       &       B7      &       $        <        28.62
         $       &       $        <       -7.18  $        \\ 
 104308       &        \newrule       21 07 51.2      &       -54 12 59
 &          326.6        &       $        <          43.6        $       &
    66.4         &       1       &       A5      &       $        <        29.26
         $       &       $        <       -5.12  $        \\ 
 118121       &        \newrule       23 57 35.0      &       -64 17 53
 &          283.0        &       $        <          81.0        $       &
    48.7         &       1       &       A1      &       $        <        29.26
         $       &       $        <       -5.50  $        \\ \hline
\multicolumn{10}{c}{\em Improbable Tuc members} \newrule \\ \hline
 459          &        \newrule       00 05 28.3      &       -61 13 32
 &          156.1        &       $        <          33.7        $       &
    53.8         &       1       &       G5      &       $        <        28.97
         $       &       $        <       -4.37  $        \\ 
 1399         &        \newrule       00 17 30.3      &       -59 57 04
 &          136.7        &       $        <          43.1        $       &
    44.3         &       1       &       M0      &       $        <        28.91
         $       &       $        <       -2.79  $        \\ 
 93096        &        \newrule       18 57 56.6      &       -44 58 06
 &          149.9        &       $        <          27.0        $       &
    64.4         &       2       &       G8    &       $        <        29.03
         $       &       $        <       -3.99  $        \\ 
 94051        &        \newrule       19 08 51.1      &       -54 02 17
 &          172.1        &       $        <          11.8        $       &
    68.4         &       1       &       G0      &       $        <        28.72
         $       &       $        <       -4.88  $        \\ 
 94858        &        \newrule       19 18 09.8      &       -53 23 13
 &          142.8        &       $        <           1.0        $       &
    45.5         &       1       &       F7      &       $        <        27.28
         $       &       $        <       -6.90  $        \\ 
 94997        &        \newrule       19 19 49.6      &       -53 43 13
 &          101.1        &       $        <          41.4        $       &
    59.9         &       1       &       M3      &       $        <        29.15
         $       &       $        <       -2.18  $        \\ 
 95302        &        \newrule       19 23 20.5      &       -50 41 20
 &          101.9        &       $        <          14.7        $       &
    75.5         &       1       &       G6    &       $        <        28.90
         $       &       $        <       -4.62  $        \\ 
 97705        &        \newrule       19 51 23.6      &       -58 30 34
 &          140.6        &       $        <           1.2        $       &
    67.8         &       1       &       F8    &       $        <        27.73
         $       &       $        <       -6.15  $        \\ 
 101636       &        \newrule       20 36 02.3      &       -54 56 28
 &          270.5        &       $        <          29.2        $       &
    65.9         &       1       &       G0      &       $        <        29.08
         $       &       $        <       -4.52  $        \\ 
 101844       &        \newrule       20 38 19.4      &       -55 36 19
 &          258.7        &       $        <          36.2        $       &
    32.0         &       1       &       K4      &       $        <        28.55
         $       &       $        <       -3.18  $        \\ 
 104256       &        \newrule       21 07 17.5      &       -57 01 55
 &           90.3        &       $        <          37.7        $       &
    53.5         &       1       &       K1      &       $        <        29.01
         $       &       $        <       -4.26  $        \\ 
 107806       &        \newrule       21 50 23.7      &       -58 18 17
 &          346.6        &       $        <          33.8        $       &
    40.8         &       1       &       G6      &       $        <        28.73
         $       &       $        <       -4.70  $        \\ 
 109612       &        \newrule       22 12 16.8      &       -54 58 40
 &          346.7        &       $        <          37.2        $       &
    49.0         &       1       &       K3      &       $        <        28.93
         $       &       $        <       -3.87  $        \\ 
 114236       &        \newrule       23 08 12.2      &       -63 37 41
 &          445.3        &       $        <           0.2        $       &
    56.7         &       1       &       G3      &       $        <        26.73
         $       &       $        <       -6.82  $        \\ \hline
\end{tabular}
\end{center}
\end{table*}

\section{The age of the Tuc association (Comparison with other star-forming regions)}\label{sect:xldfs}

In this section the XLDF of the Tucanae candidates is studied. 
The X-ray emission declines with stellar age. Therefore, 
a comparison to XLDFs of other nearby regions of star formation
should provide insight into the evolutionary stage of this association.
In the following we describe the comparison samples.

\subsection{TW\,Hydrae}\label{subsect:twa}

Until the discovery of the Tucanae association, the TW\,Hydrae association
was the only recognized nearby young stellar association far from molecular
clouds.
Strong X-ray emission is considered as (one of many) indicators for 
membership to the association since it was first studied in X-rays 
by \citey{Kastner97.1}. 
In subsequent studies which have added more stars to the group X-ray 
luminosities for the new members have been presented (see e.g. 
\cite{Jensen98.1}, \cite{Hoff98.1}, \cite{Webb99.1}, \cite{Sterzik99.1}, \cite{Hoff00.1}).
However, a systematic study of the X-ray properties of the whole sample has
not been performed, and no XLDFs have been examined so far. In view of the
similarity of the Tucanae and TW\,Hydrae regions, 
particularly the close and similar distance, 
it is intimating to compare their X-ray characteristics. 

Fourteen TTS systems are known so far in the TW Hya region
(see \cite{Rucinski83.1}, \cite{delaReza89.1}, 
\cite{Gregorio-Hetem92.1}, \cite{Kastner97.1},
\cite{Jensen98.1}, \cite{Webb99.1}, \cite{Sterzik99.1},
\cite{Hoff00.1}), all listed in Table~\ref{tab:twa}.
The multiplicities of these objects are given in \citey{Webb99.1},
\citey{Sterzik99.1}, and \citey{Hoff00.1}.
The possible spectroscopic binaries listed by \citey{Webb99.1}
have been confirmed as such (Torres et al., in prep.).
The recently detected faint object next to TWA-7
(\cite{Neuhaeuser00.1}) has most recently been found to 
be a background K dwarf according to an H-band spectrum
taken with ISAAC at the VLT (\cite{Neuhaeuser00.2})
so that we regard TWA-7 as single.

We have cross-correlated the list of TW Hydrae stars with
the RASS BSC. All but one
star (GSC\,7739\,2180) can be identified with an X-ray source at less than 
$40^{\prime\prime}$ from the optical position. 
In order to obtain the X-ray properties of GSC\,7739\,2180 
which had no counterpart in the BSC we have checked the raw RASS data, 
and found that the star is located in a region of low exposure. We have
detected an X-ray source with 10.7\,cts at the optical position
(RX\,J1121.1-3845). Thus, 
all TW\,Hydrae stars are X-ray emitters.
Their X-ray characteristics
are listed in Table~\ref{tab:twa}. We give the
designation of the star (column~1), the X-ray position (column~2~and~3), 
the distance between optical and X-ray position $\Delta$ (column~4), 
the exposure time (column~5), the broad band count rate (column~6), the
multiplicity (column~7), X-ray luminosity (column~8) and hardness ratios 
(column~9~and~10). As for the Tucanae stars, the luminosity of multiples 
has been divided by the number of components.
Our X-ray data agree well with previously published {\em ROSAT} data
for TWA stars\footnote{For HR\,4796, a pair comprised of an A type and an M
type star with separation of just $7.6^{\prime\prime}$, 
an elongated X-ray source has
been detected in the HRI pointed observation. We can therefore distribute
the photons detected here in the RASS equally among the two components, as
in the other binaries. Note that \citey{Huelamo00.1} give a different HRI
X-ray luminosity for HR\,4796 than \citey{Jura98.1}, because
\citey{Jura98.1} converted the HRI X-ray photons using the PSPC energy
conversion factor, while \citey{Huelamo00.1} used the correct HRI conversion.}.
\begin{table*}
\begin{center}
\caption{RASS X-ray data of stars in the TW\,Hydrae association. X-ray data are
derived from the RASS Bright Source Catalog with the exception of 
RX\,J1121.1-3845 (GSC\,7739\,2180), which
is in a region of low exposure and therefore has no entry in the BSC. For
this star we have extracted and analyzed the RASS raw data. The distances
of the stars are computed from the {\em Hipparcos} parallax when available,
otherwise we adopt a value of 55\,pc, the mean of the {\em Hipparcos}
distances measured for four TW\,Hya members. Multiples are
listed only once. The meaning of the columns is the same as in Table~\ref{tab:all_sol}.}
\label{tab:twa}
\begin{tabular}{lrrrrrrrrr}\hline
Designation & \multicolumn{2}{c}{BSC X-ray position} & \multicolumn{1}{c}{$\Delta$} & \multicolumn{1}{c}{Expo} & \multicolumn{1}{c}{Rate} & Multi & \multicolumn{1}{c}{$L_{\rm x}$} & \multicolumn{1}{c}{$HR1$} & \multicolumn{1}{c}{$HR2$} \\ 
              & \multicolumn{1}{c}{[R.A.: h\,m\,s]} &              \multicolumn{1}{c}{[Dec.: d\,m\,s]} &              \multicolumn{1}{c}{[$^{\prime\prime}$]} &              \multicolumn{1}{c}{[s]} & \multicolumn{1}{c}{[cts/s]} & &  \multicolumn{1}{c}{[$10^{29}\,{\rm erg/s}$]} & & \\ \hline
 TWA-6           &        \newrule       10 18 28.8      &       -31 50 02 &          1.4  &          336  &       $             0.29      \pm           0.03      $       &       2       &       $           3.6         \pm         0.9         $       &       $        -0.29  \pm       0.10  $       &       $   0.30  \pm       0.17  $        \\ 
 TWA-7           &        \newrule       10 42 30.3      &       -33 40 17 &          2.5  &          370  &       $             0.32      \pm           0.03      $       &       1       &       $           9.2         \pm         1.0         $       &       $        -0.08  \pm       0.09  $       &       $   0.05  \pm       0.14  $        \\ 
 TW Hya$^a$          &        \newrule       11 01 51.9      &       -34 42 17 &          5.2  &          337  &       $             0.57      \pm           0.04      $       &       1       &       $          24.8         \pm         7.0         $       &       $         0.58  \pm       0.06  $       &       $  -0.12  \pm       0.08  $        \\ 
 CoD-29 8887   &        \newrule       11 09 13.9      &       -30 01 39 &          7.8  &          325  &       $             0.34      \pm           0.03      $       &       2       &       $           4.4         \pm         0.9         $       &       $        -0.22  \pm       0.09  $       &       $  -0.02  \pm       0.15  $        \\ 
 RX\,J1109.7-3907      &        \newrule       11 09 40.1      &       -39 06 48 &          9.8  &          367  &       $             0.12      \pm           0.02      $       &       1       &       $           3.8         \pm         0.7         $       &       $         0.15  \pm       0.16  $       &       $   0.08  \pm       0.21  $        \\ 
 Hen 3-600     &        \newrule       11 10 28.0      &       -37 32 07 &         11.1  &          336  &       $             0.28      \pm           0.03      $       &       3       &       $           2.8         \pm         0.3         $       &       $        -0.01  \pm       0.11  $       &       $  -0.06  \pm       0.16  $        \\ 
 RX\,J1121.1-3845$^b$   &        \newrule       11 21 05.2      &       -38 45 27 &          0.0  &          100  &       $             0.11      \pm           0.04      $       &       1       &       $           3.3         \pm         1.2         $       &       $         0.05  \pm       0.39  $       &       $   0.02  \pm       0.52  $        \\ 
 RX\,J1121.3-3447   &        \newrule       11 21 16.8      &       -34 46 43 &          3.9  &          473  &       $             0.43      \pm           0.03      $       &       2       &       $           6.1         \pm         0.6         $       &       $        -0.08  \pm       0.07  $       &       $   0.03  \pm       0.11  $        \\ 
 HD 98800$^a$        &        \newrule       11 22 05.2      &       -24 46 40 &          8.3  &          330  &       $             0.66      \pm           0.05      $       &       4       &       $           3.7         \pm         0.3         $       &       $         0.06  \pm       0.06  $       &       $  -0.02  \pm       0.10  $        \\ 
 CoD-33 7795   &        \newrule       11 31 55.5      &       -34 36 27 &          5.6  &          122  &       $             0.66      \pm           0.08      $       &       3       &       $           5.3         \pm         1.4         $       &       $        -0.31  \pm       0.10  $       &       $   0.29  \pm       0.18  $        \\ 
 TWA-8         &        \newrule       11 32 41.5      &       -26 51 55 &          2.7  &          307  &       $             0.32      \pm           0.04      $       &       2       &       $           4.9         \pm         0.5         $       &       $         0.03  \pm       0.10  $       &       $  -0.06  \pm       0.15  $        \\ 
 CoD-36 7429$^a$   &        \newrule       11 48 24.2      &       -37 28 49 &         11.4  &          122  &       $             0.34      \pm           0.06      $       &       2       &       $           3.7         \pm         0.9         $       &       $        -0.23  \pm       0.16  $       &       $  -0.18  \pm       0.26  $        \\ 
 TWA-10          &        \newrule       12 35 04.3      &       -41 36 39 &          9.5  &          341  &       $             0.12      \pm           0.02      $       &       1       &       $           3.8         \pm         0.7         $       &       $         0.08  \pm       0.16  $       &       $  -0.29  \pm       0.20  $        \\ 

 HR 4796\,A$^{a,c}$       &        \newrule       12 36 01.0      &       -39 52 10 &         15.8  &          284  &       $             0.11      \pm           0.02      $       &       2       &       $           2.4         \pm         0.6         $       &       $        -0.24  \pm       0.17  $       &       $   0.53  \pm       0.19  $        \\ 
 HR 4796\,B$^{a,c}$       &        \newrule       12 36 00.6      &       -39 52 16 &         14.9  &  &   &       2       &       $           2.1         \pm         0.6         $       &     &         \\ \hline
\end{tabular}
$^a$ Distance from the {\em Hipparcos} parallax, 
$^b$ Position from the RASS raw data.
$^c$ Components A and B have different $L_{\rm x}$ because of the
difference in the $ECF$ between early and late spectral types (HR\,4796\,A
is an A0 star, and HR\,4796\,B an M2 star; see also Sect.~\ref{subsect:pspc}).
\end{center}
\end{table*}

\subsection{Taurus-Auriga}\label{subsect:taurus}

Taurus-Auriga is one of the nearest ($d=140$\,pc; \cite{Elias78.1}, 
\cite{Wichmann98.1}) and best studied regions of star formation. The
region is particularly rich in late-type PMS stars.
In an analysis of RASS data \citey{Neuhaeuser95.1} found that the subclass
of weak-line TTS (i.e. TTS with weak H$\alpha$ emission lines) in 
Taurus-Auriga are X-ray brighter than classical TTS (which are defined by
equivalent widths of H$\alpha > 10\,{\rm \AA}$). 
Because the Tucanae members appear to be somewhat evolved, they are
probably all weak H$\alpha$ emitters (naked weak-line post-TTS on
radiative tracks near the ZAMS). Therefore, for our comparison we select 
only the weak-line TTS from Taurus-Auriga. We also note, that
the classical TTS in Taurus-Auriga appear to be less X-ray bright than
the wTTS although being younger (\cite{Neuhaeuser95.1}, 
Stelzer et al. in prep). This may be due to magnetic star-disk coupling 
during the cTTS phase 
which prevents spin-up, and restricts the dynamo efficiency.
The X-ray luminosity, therefore, shows a peak at the stage of the wTTS, and 
the decrease in X-ray luminosity with age sets in only after
the disk is dissipated.

We have re-computed 
the XLDF for the RASS data of Taurus-Auriga weak-line TTS including also
stars which have been discovered since the study of \citey{Neuhaeuser95.1}.
These newly discovered stars are listed in \citey{Koenig00.1}. Most
of them were not detected during the RASS.
We do not include all those TTS, which were originally discovered by
{\em ROSAT}, in order to avoid a bias towards X-ray bright TTS.

\subsection{IC\,2602}\label{subsect:ic2602}

IC\,2602 is a 30\,Myr (\cite{Mermilliod81.1}) 
old open cluster whose stars are about to 
reach the main-sequence. With a distance of $\sim 150\,{\rm pc}$ 
(see \cite{Whiteoak61.1}) the
cluster is relatively nearby. The X-ray emission from IC\,2602 was
studied by \citey{Randich95.1} who have analyzed pointed PSPC observations
and selected probable cluster members on the basis of the photometry 
of the optical counterparts. For the comparison of the XLDFs we have
made use of Table~4 in \citey{Randich95.1} (see Sect.~\ref{subsect:xldfs}).

\subsection{Pleiades}\label{subsect:pleiades}

Due to their relatively small distance (116\,pc; \cite{Mermilliod97.1}) 
the $100\,{\rm Myr}$ old Pleiades cluster 
has often been used in comparative evolutionary studies.
Detailed investigations of the X-ray emission from the Pleiades have been 
presented e.g. by \citey{Stauffer94.1}, \citey{Micela96.1}, and
\citey{Micela99.1} based on observations performed by the {\em Einstein}
and {\em ROSAT} satellites. A study of the full set of archived {\em ROSAT}
observations will be presented in a later publication (Stelzer et
al., in prep). Here, we anticipate the XLDF composed of all Pleiades stars
which have been in the field of view of any pointed PSPC observation. Upper
limits for non-detections are included. A detailed description of the
data analysis will be given in the subsequent paper.

\subsection{X-ray Luminosity Functions}\label{subsect:xldfs}

To compute the XLDFs for the different stellar associations we 
have used the ASURV statistics package (see \cite{Feigelson85.1}) which
ensures a proper treatment of censored data points, i.e upper limits for 
undetected sources.
For unresolved multiples we have assumed that all components emit X-rays at
the same level. The observed $L_{\rm x}$ has, therefore, been divided by
the number of components, and all components are considered in the XLDF.

\begin{figure}
\begin{center}
\resizebox{9cm}{!}{\includegraphics{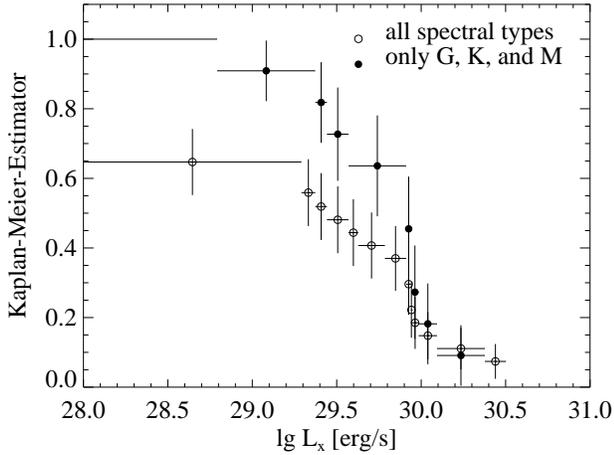}}
\caption{RASS XLDF for stars from \protect\citey{Zuckerman00.1}: (a) probable
members with any spectral type, (b) probable members with late spectral type. See text for a description of the samples.}
\label{fig:ldf_tuc}
\end{center}
\end{figure}

In Fig.~\ref{fig:ldf_tuc} we display the RASS XLDF of the group of probable
members of the Tucanae association. The subsample of late-type stars is
also shown. For these stars the XLDF is somewhat steeper indicating that
the early-type stars are the weaker X-ray emitters. Five of the
early-type stars in the Tucanae sample 
have spectral type A. For A type stars no
mechanism producing X-rays is known, consistent with our finding that all A
stars in Tucanae are undetected in the RASS. O and B stars can generate X-rays
in shocks associated with their strong winds. None of the B stars
from our sample is detected in the RASS, but one B star is detected in a
PSPC pointing. For late-type stars with 
convective envelopes of substantial depth 
(starting from $\sim$ F5; \cite{Walter83.1}) the X-ray emission is
thought to be related to magnetic dynamo activity. 
All stars in Taurus-Auriga are TTS, i.e. late-type PMS stars.
Therefore, to obtain
homogeneous samples for the comparison of XLDFs with other star-forming
regions only the G, K, and M stars among the Tucanae members
should be retained.
Due to the non-detection of stars earlier than spectral type F in the
Tucanae sample, 
reducing the sample to G, K, and M stars mainly effects the
number of upper limits involved in the XLDF.

We have reduced the other samples in the same way to G, K, and M
members. In TW\,Hydrae most stars have very late spectral 
types. Only one star, HR\,4796A (spectral type A0), had to be excluded. 
All TTS in Taurus have spectral types G and later.
The sample of IC\,2602 is composed of all stars from Table~4
in \citey{Randich95.1} which are labeled `photometric members' (flags `Y'
or `Y?') and have $B-V > 0.6$. In this table the authors give 
$\lg{L_{\rm x}}$ for all {\em ROSAT} PSPC X-ray sources in the cluster 
with optical counterparts. To treat the multiples among these
stars consistently with the other stellar groups 
we have made use of the Open Cluster Data Base
compiled by C. Prosser and colleagues (available at ftp://cfa-ftp.harvard.edu/pub/stauffer/clusters). The
multiplicities given in the Open Cluster Data Base are used to derive
$L_{\rm x}$ of the individual components as described in 
Sect.~\ref{subsect:pspc}.
The membership list in the Pleiades is based on the entries of
the Open Cluster Data Base. For the XLDF computed for comparison with
Tucanae we restrict this sample to G, K, and
M type stars which have been observed in any pointed PSPC
observation (see Stelzer et al., in prep. for more details).

\begin{figure}
\begin{center}
\resizebox{9cm}{!}{\includegraphics{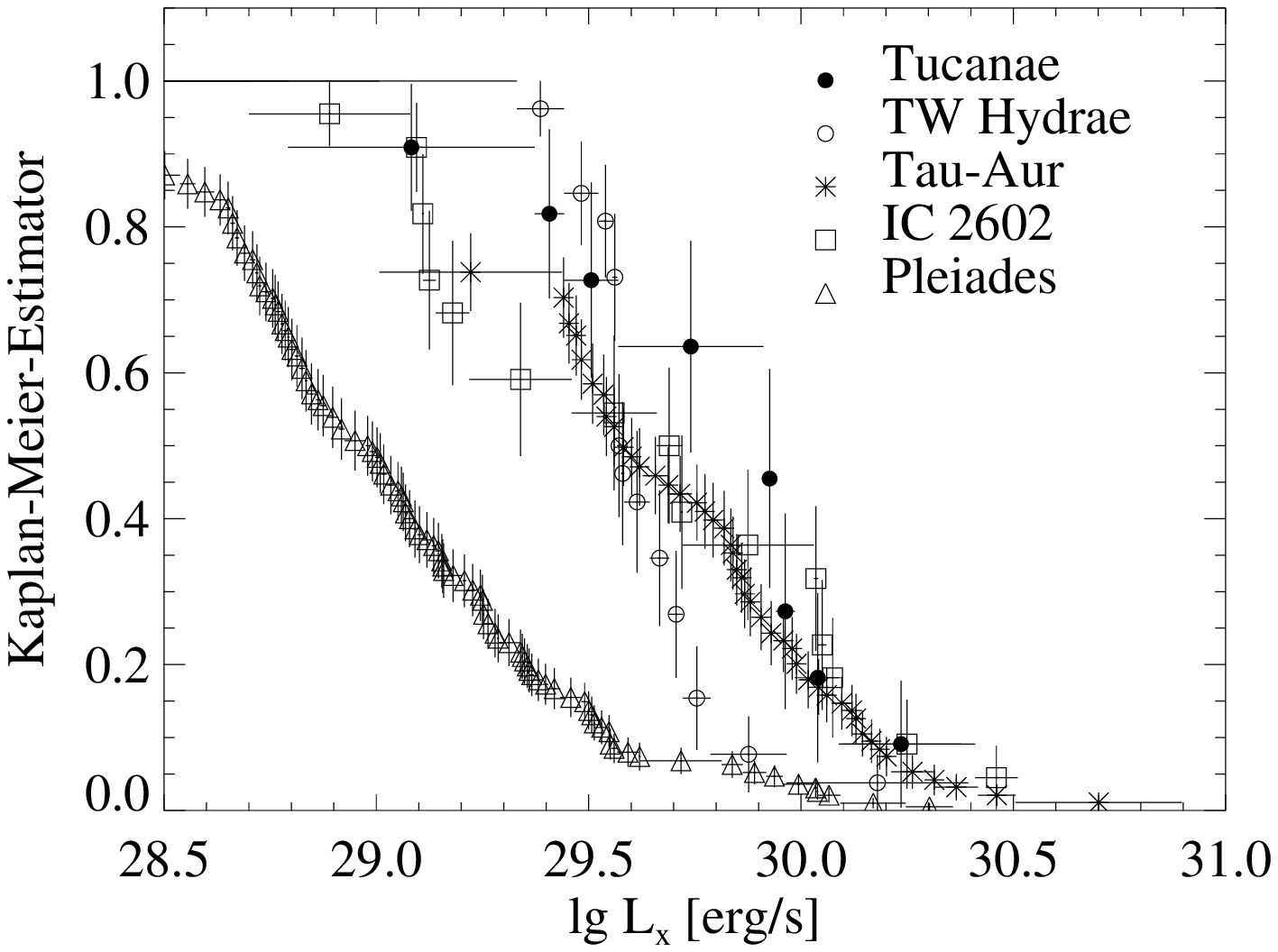}}
\caption{Comparison between XLDFs of late-type stars (spectral types 
G, K, and M) for different star forming regions as
observed by {\em ROSAT}: {\em filled circles} - Tucanae, {\em open circles} -
TW\,Hydrae, {\em crosses} - Taurus-Auriga, {\em squares} - IC\,2602, and
{\em triangles} - Pleiades. See text for a description of the samples.}
\label{fig:ldf_regions}
\end{center}
\end{figure}

The Kaplan-Meier Estimator for the late-type stars of 
all stellar groups introduced in the previous 
subsections is shown in Fig.~\ref{fig:ldf_regions}.
All distributions except that of the Pleiades are remarkably 
similar. Particularly, 
the XLDF of the 30\,Myr old IC\,2602 cluster and the PMS regions occupy the
same region in the diagram.
The distribution for TW\,Hydrae shows the steepest slope, i.e. smallest
spread of luminosities. The narrow luminosity distribution of the
TW\,Hydrae sample might be explained by two effects: 
(i) Since only for four systems
the parallax has been measured, we have adopted a mean distance of 55\,pc
for the remaining stars. Therefore the real spread in distance is probably
underestimated. And (ii) the spectral
type distribution is very homogeneous in TW\,Hydra. 
Most stars have spectral types late K or M, while for the other
samples the spread in spectral types is larger (a substantial number of G 
and early K stars enter the distributions).
We note, that the distribution of the Pleiades and IC\,2602 have
been derived from pointed data, while the XLDF of Tucanae, TW\,Hydrae and
Taurus-Auriga are obtained from RASS observations. However, in the displayed
luminosity range the lower sensitivity limit of the RASS should not
play a role, and all distributions should be complete.

The general coincidence of the shape and location of the XLDF of Tucanae,
TW\,Hydrae, Taurus-Auriga, and IC\,2602 suggests that the stars in the
Tucanae association are young ($10$ to $30$\,Myr). Certainly, their age is
well below that of the Pleiades ($10^8\,{\rm yrs}$) whose XLDF is clearly 
shifted to lower luminosities. This is also manifest in 
Table~\ref{tab:meanlx} where we give the
mean and median of the X-ray luminosity for all examined samples.
\begin{table}
\begin{center}
\caption{Mean and median of the X-ray luminosity $L_{\rm x}$ for the
stellar samples compared in Fig.~\protect\ref{fig:ldf_regions} computed
with ASURV. Results for all stars (regardless of spectral type), and for
the samples restricted to G, K, and M stars are given. The number of stars
in each sample and number of upper limits among them are listed in columns
`Size' and `(u.l.)'.}
\label{tab:meanlx}
\begin{tabular}{llrrrr} \hline
Sp.Type & Sample & Size & (u.l.) & \multicolumn{2}{c}{$\lg{L_{\rm x}}\,[{\rm erg/s}]$} \\ 
& & & & \multicolumn{1}{c}{Mean} & \multicolumn{1}{c}{Median} \\ \hline
All & Tucanae & 27 & (10) & $29.19 \pm 0.18$ & $29.41$ \\
All & TW\,Hydrae & 27 & (0) & $29.63 \pm 0.04$ & $29.57$ \\ \hline
G,K,M & Tucanae & 11 & (1) & $29.76 \pm 0.13$ & $29.83$ \\
G,K,M & TW\,Hydrae & 26 & (0) & $29.64 \pm 0.04$ & $29.57$ \\
G,K,M & Tau-Aur & 95 & (35) & $29.61 \pm 0.05$ & $29.57$ \\ 
G,K,M & IC\,2602 & 22 & (0) & $29.61 \pm 0.11$ & $29.66$ \\
G,K,M & Pleiades & 192 & (66) & $29.00 \pm 0.04$ & $28.97$ \\ \hline
\end{tabular}
\end{center}
\end{table}
The values presented in Table~\ref{tab:meanlx} have been derived with
ASURV, i.e. upper limits have been taken account of. The weakly X-ray
emitting stars of early spectral type in Tucanae reduce 
$\langle \lg{L_{\rm x}} \rangle$ significantly. All values for 
$\langle \lg{L_{\rm x}} \rangle$ of late-type stars 
except the Pleiades are compatible with
each other within their uncertainties.

\begin{figure}
\begin{center}
\resizebox{9cm}{!}{\includegraphics{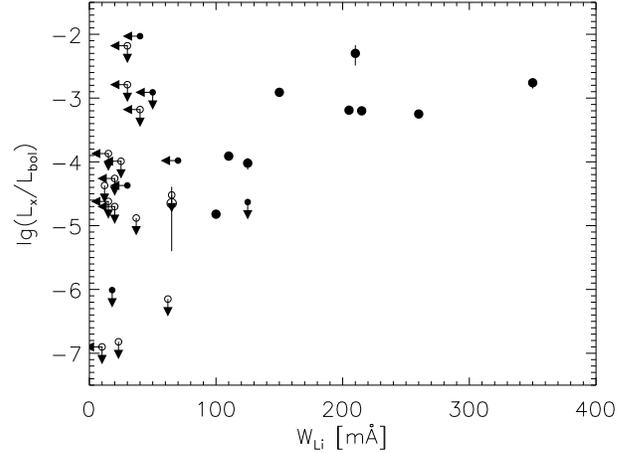}}
\caption{Correlation between the X-ray to bolometric luminosity ratio and 
the equivalent width of Lithium for stars of the Tucanae
association: Filled symbols are probable members (according to 
\protect\cite{Zuckerman00.1}). Improbable members are represented by open 
symbols. Arrows indicate upper limits.}
\label{fig:lx_wli}
\end{center}
\end{figure}

The mean ratio of the logarithm 
of the X-ray to bolometric luminosity for the RASS detected
Tucanae members (Table~1a of \cite{Zuckerman00.1}) is 
$-3.47 \pm 0.19$, which is typical for late-type stars generally
characterized by $\lg{(L_{\rm x}/L_{\rm bol})} \simeq -3$. Four Tucanae stars
display an $L_{\rm x}/L_{\rm bol}$ value slightly higher than this
saturation limit. But note, that two of these show strong variability
(see lightcurves in Fig.~\ref{fig:lcs_rass}) 
which may have led to an overestimation of the quiescent X-ray emission.
For the probable members $L_{\rm x}/L_{\rm bol}$ correlates well with the 
equivalent width of Li\,I, a commonly accepted age indicator. 
The correlation is shown in Fig.~\ref{fig:lx_wli} (filled circles). 
The stars labeled as `improbable' members by
\citey{Zuckerman00.1} do not follow the $L_{\rm x}/L_{\rm bol}-W_{\rm Li}$ 
relation (open circles).
In most stars from this group neither the Li I line is detected nor are
they known to be X-ray emitters consistent with them not being part of the 
association.

\section{Variability}\label{sect:variability}

In order to investigate whether the X-ray emission of the Tucanae stars is
variable we have generated 
lightcurves for all detected sources from the arrival time information
of the photons counted within the source extraction radius. 
The background counts were extracted from the background map as
described in Sect.\ref{subsect:pspc}. An alternative method of background
acquisition consists of constructing a separate background lightcurve at a 
source free position, and subtracting it from the source lightcurve. This
procedure is sensitive to local variations in the background. We have
applied both methods to the Tucanae stars and found no significant 
differences. The count rates have been corrected for vignetting,
i.e. effects due to detector off-axis position and support structure.

For the RASS lightcurves the photons have been binned into 5600\,s intervals 
(corresponding to the
duration of one satellite orbit around the Earth). Note, however, that
the actual exposure times in individual bins range between $\sim$ 10 -
30\,s.  For pointed observations we use a binsize of 400\,s. This
interval corresponds to the wobbling motion of the telescope. The wobble is
performed to ensure that no source remains hidden behind the entry 
window's support structure, and may produce apparent variability in a
lightcurve if the binning is smaller than that period.

Most of the lightcurves show strong variability. To examine the variations
in a quantitative way 
we have computed the relative luminosity change between the
bin with maximum ($C_{\rm max}$) and minimum count rate ($C_{\rm min}$) 
for each star, i.e.
\begin{equation}
\Delta C_{\rm x} = \frac{C_{\rm max} - C_{\rm min}}{\frac{1}{2}\,(C_{\rm max} +
C_{\rm min})}
\label{eq:deltaC}
\end{equation}
and its 1\,$\sigma$ uncertainty from the errors of the respective count
rates. 
\begin{figure*}
\begin{center}
\resizebox{18cm}{!}{\includegraphics{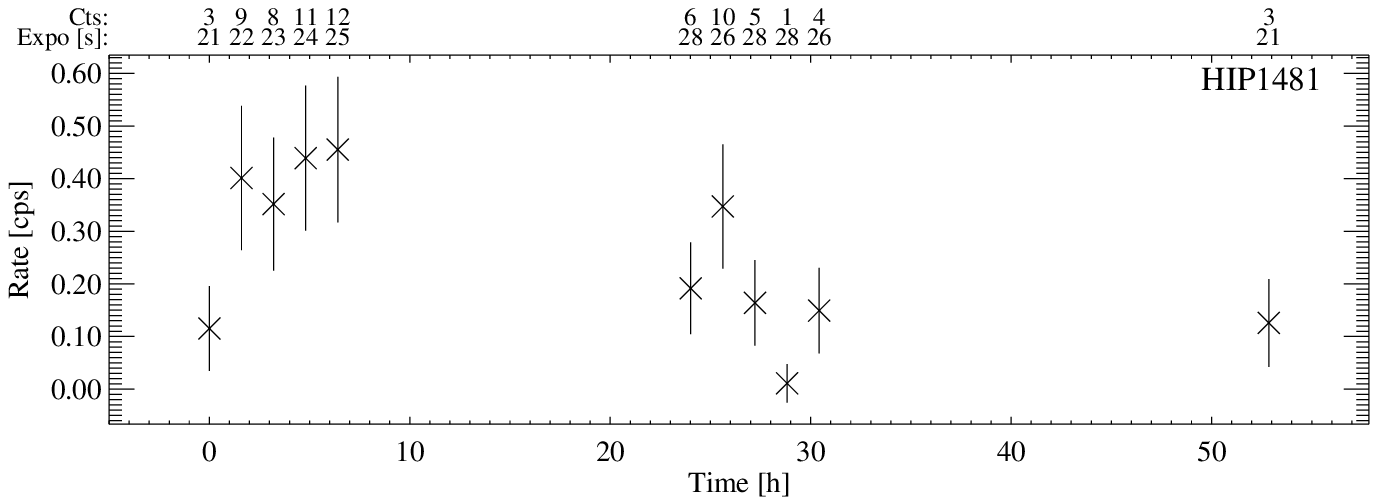}}
\resizebox{18cm}{!}{\includegraphics{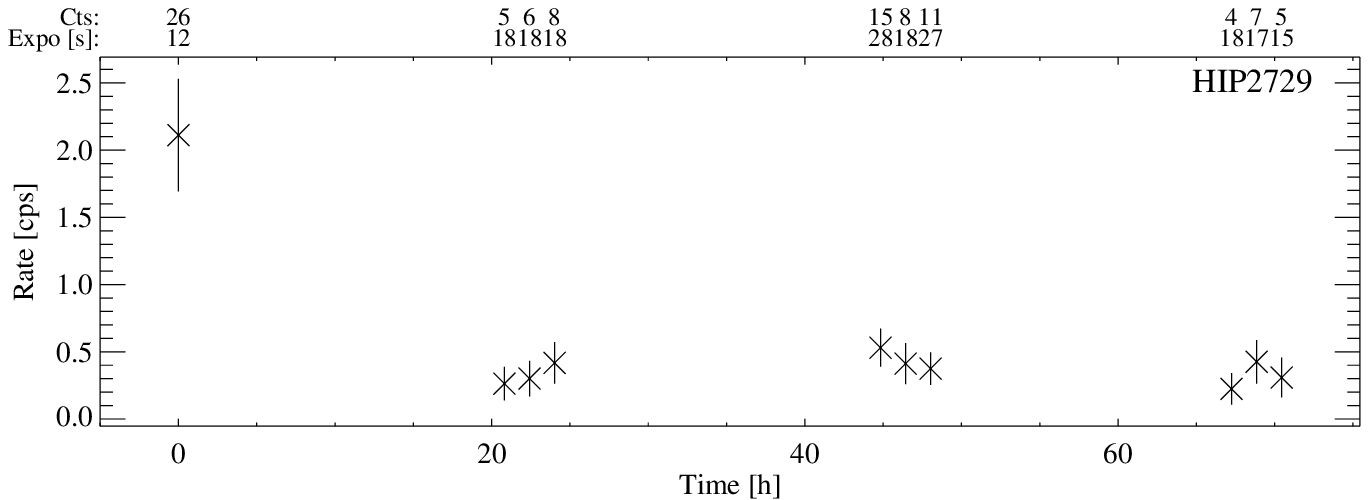}}
\resizebox{18cm}{!}{\includegraphics{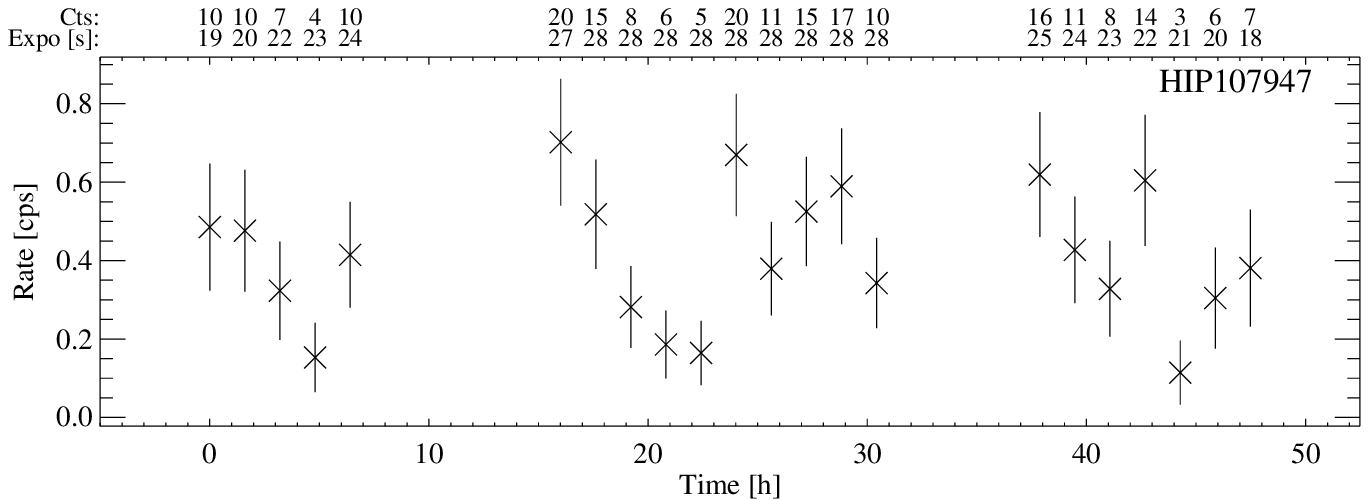}}
\resizebox{18cm}{!}{\includegraphics{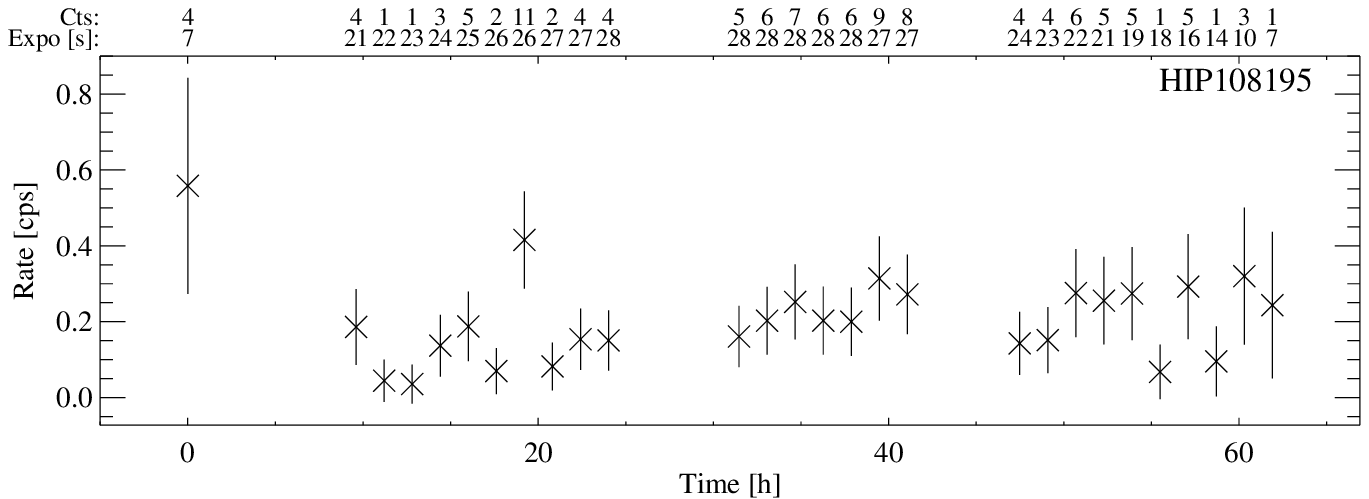}}
\caption{RASS lightcurves of variable stars, i.e. stars for
which the relative change in count rate $\Delta C_{\rm x}$ within the
observation is significant
at $\geq 3\,\sigma$. Shown are 1\,$\sigma$ uncertainties. Arrows indicate
the background for scans without source counts. The labels on top of each
panel give the number of counts and exposure time
(in s) for each scan.}
\label{fig:lcs_rass}
\end{center}
\end{figure*}
\addtocounter{figure}{-1}
\begin{figure*}
\begin{center}
\resizebox{18cm}{!}{\includegraphics{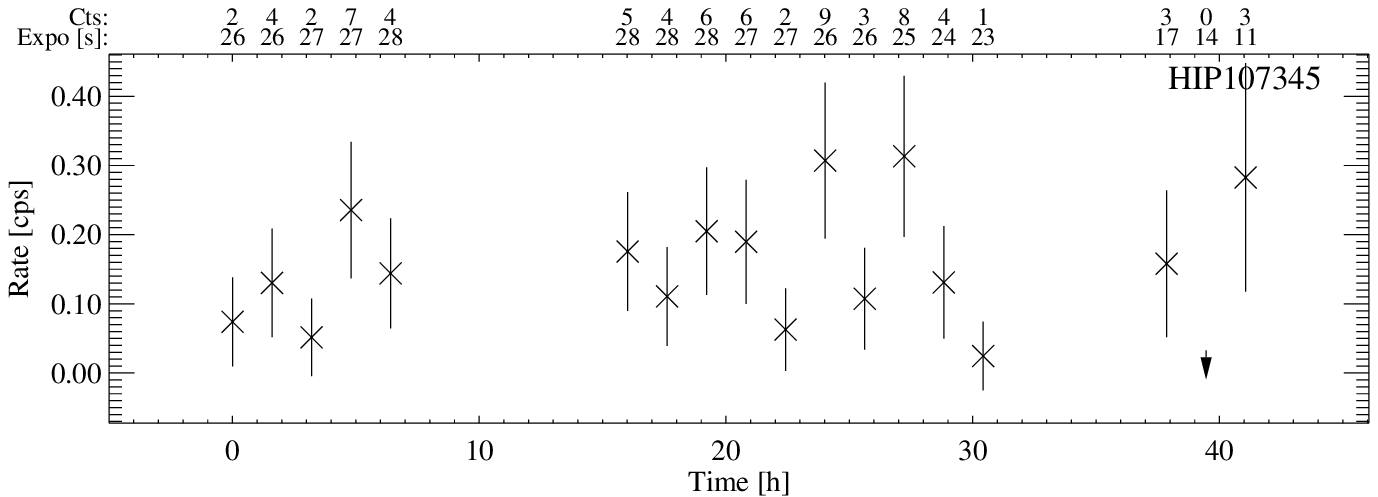}}
\resizebox{18cm}{!}{\includegraphics{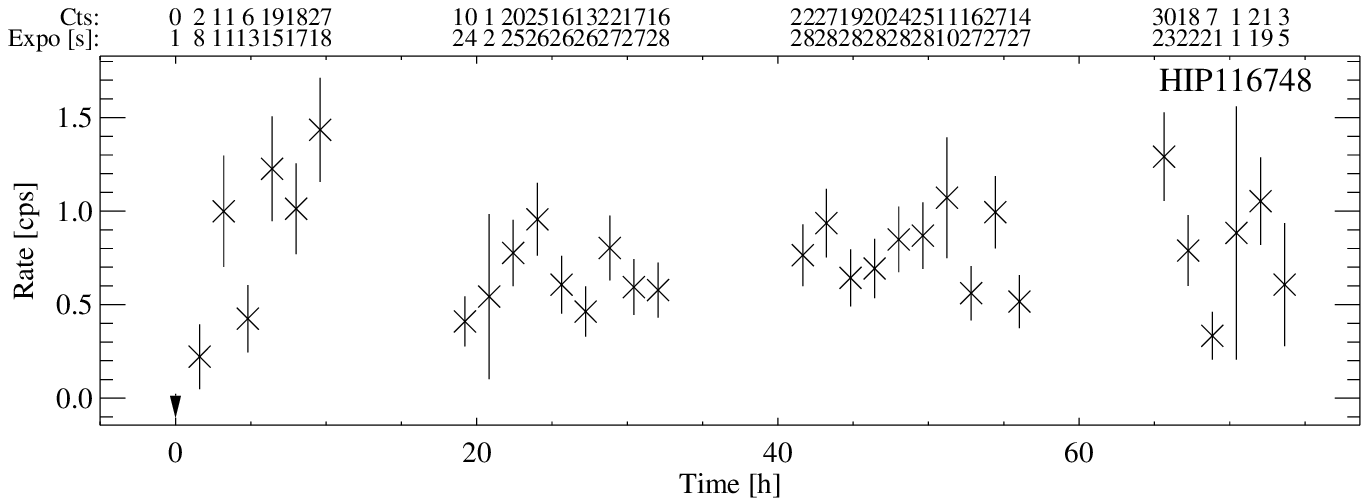}}
\resizebox{18cm}{!}{\includegraphics{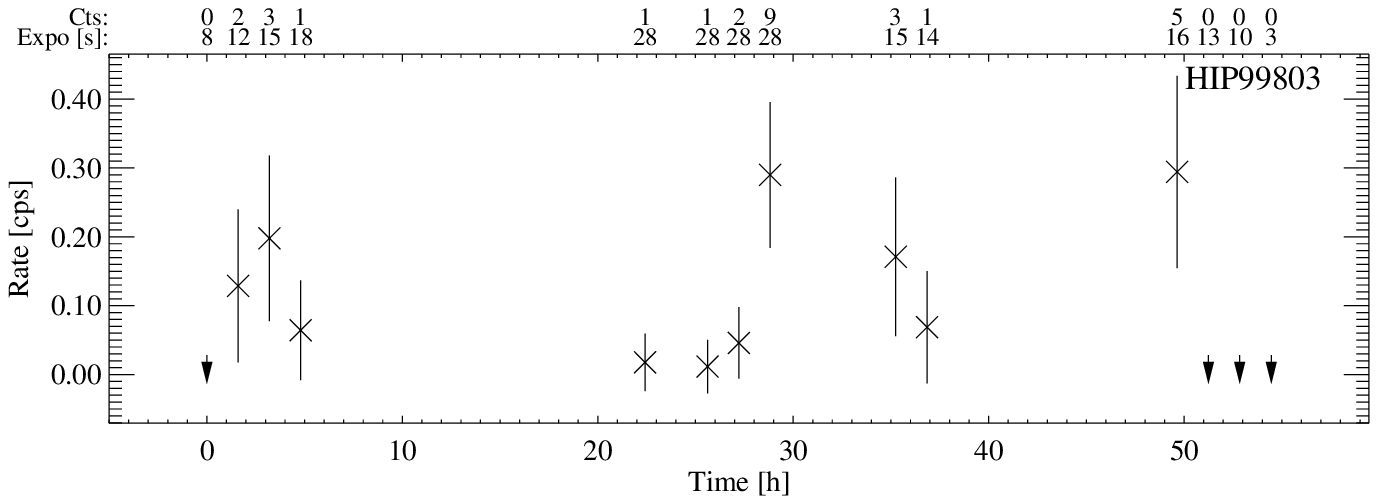}}
\resizebox{18cm}{!}{\includegraphics{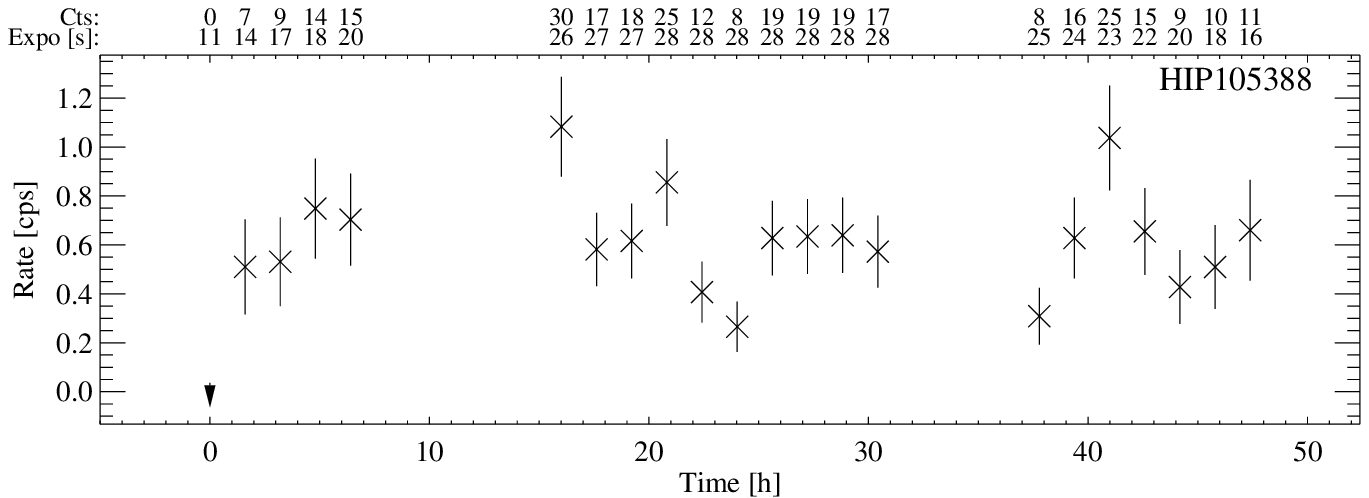}}
\caption{{\em continued}}
\end{center}
\end{figure*}
\addtocounter{figure}{-1}
\begin{figure*}
\begin{center}
\resizebox{18cm}{!}{\includegraphics{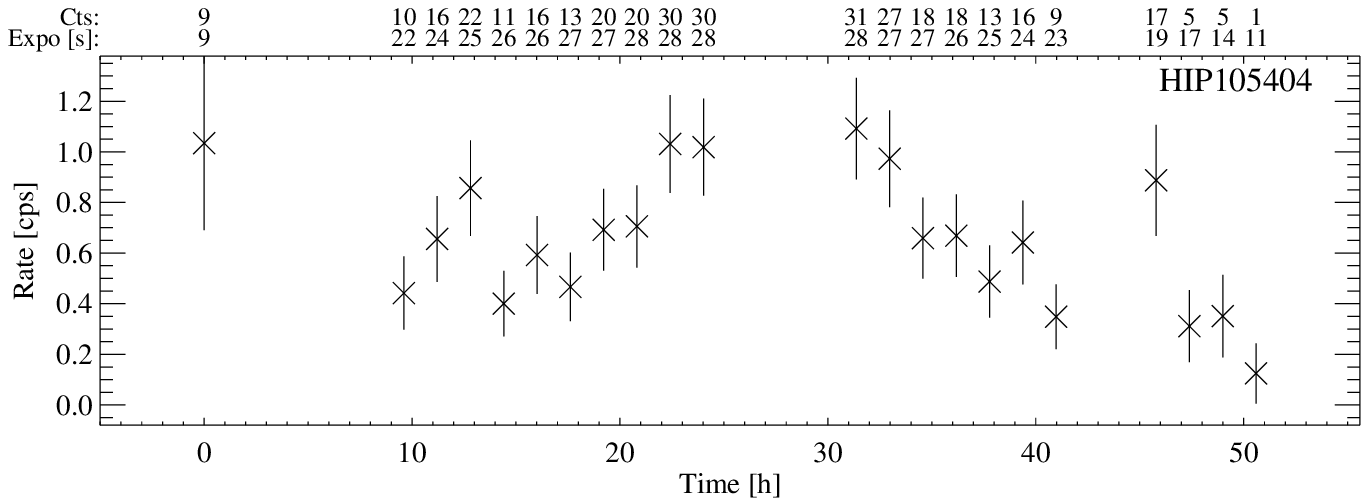}}
\resizebox{18cm}{!}{\includegraphics{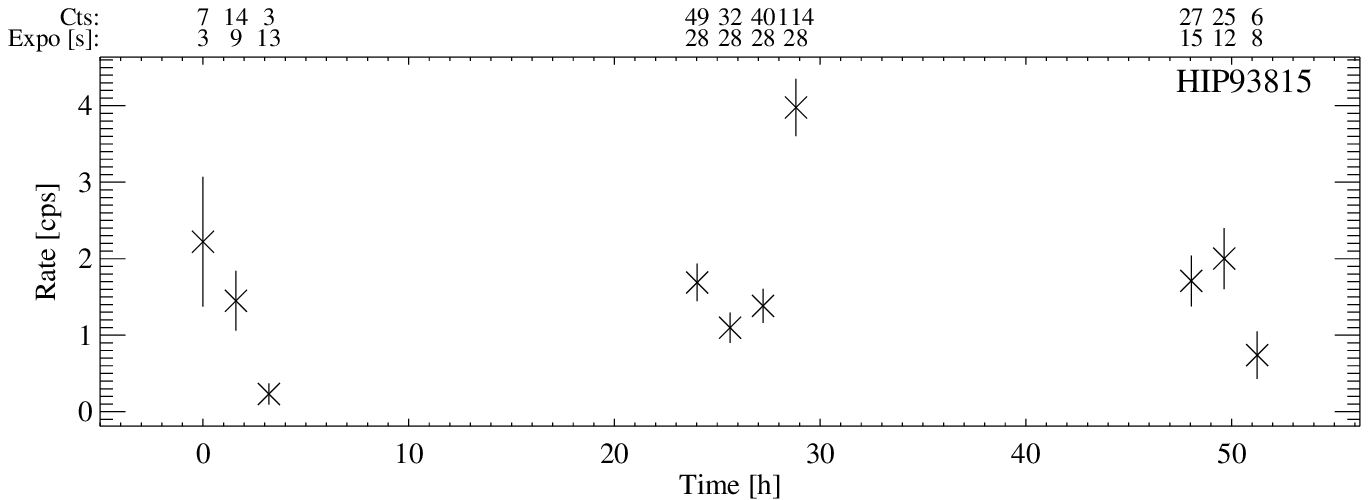}}
\caption{{\em continued}}
\end{center}
\end{figure*}
In Fig.~\ref{fig:lcs_rass} we display all 
RASS lightcurves for which the change in count rate $\Delta C_{\rm x}$
in the course of the observation amounts to 
at least 3\,$\sigma$. 
For clarity we provide the number of counts and the
exposure time in each bin on top of each lightcurve.
The variability is significant at the 3\,$\sigma$ level for 
10 of the detected Tucanae members (77\%).
The values $\Delta C_{\rm x}$ and its significance are given in 
Table~\ref{tab:deltaC} together with the minimum and maximum luminosity
observed in the lightcurve. 
\begin{table}
\begin{center}
\caption{Variability in the RASS lightcurves of the Tucanae stars measured
through the relative change in count rate $\Delta C_{\rm x}$ during the
observation (see text). Column~3 provides a measure of the significance of
the variation. All changes larger than 3\,$\sigma$ have been considered
significant, and the corresponding lightcurves are displayed in
Fig.~\ref{fig:lcs_rass}. Columns~4~and~5 are the minimum and maximum
luminosity inferred from the lightcurve using distance and $ECF$ as
described in the text.}
\label{tab:deltaC}
\begin{tabular}{lrrrr} \hline
HIP & $\Delta C_{\rm x}$ & $\frac{\Delta C_{\rm x}}{\sigma_{\Delta C_{\rm x}}}$ & $L_{\rm min}$ & $L_{\rm max}$ \\ 
& & & [$10^{29}\,{\rm erg/s}$] & [$10^{29}\,{\rm erg/s}$] \\ \hline
 1481  	 &	  \newrule 	   1.91	 &	   6.22	 &	    0.2	 &	    6.7	  \\ 
 1910  	 &	  \newrule 	   1.25	 &	   1.83	 &	    1.0	 &	    4.3	  \\ 
 2729  	 &	  \newrule 	   1.62	 &	   8.38	 &	    4.1	 &	   39.0	  \\ 
 PPM366328	 &	  \newrule 	   1.43	 &	   2.68	 &	    1.1	 &	    6.9	  \\ 
 107947	 &	  \newrule 	   1.44	 &	   3.99	 &	    2.2	 &	   13.7	  \\ 
 108195	 &	  \newrule 	   1.76	 &	   5.10	 &	    0.7	 &	   11.5	  \\ 
 107345	 &	  \newrule 	   1.71	 &	   3.08	 &	    0.4	 &	    4.9	  \\ 
 116748	 &	  \newrule 	   1.46	 &	   3.93	 &	    4.3	 &	   27.6	  \\ 
 99803 	 &	  \newrule 	   1.85	 &	   3.76	 &	    0.6	 &	   14.2	  \\ 
 105388	 &	  \newrule 	   1.21	 &	   4.46	 &	    5.1	 &	   20.8	  \\ 
 105404	 &	  \newrule 	   1.59	 &	   4.45	 &	    2.6	 &	   22.4	  \\ 
 93815 	 &	  \newrule 	   1.78	 &	  14.19	 &	    7.2	 &	  123.0	  \\ 
 92680 	 &	  \newrule 	   0.86	 &	   2.48	 &	   12.4	 &	   31.0	  \\ \hline
 103438	 &	  \newrule 	   1.80	 &	   2.63	 &	    0.2	 &	    3.0	  \\ \hline
\end{tabular}
\end{center}
\end{table}
For variable stars the amplitude of the lightcurve is displayed in 
Fig.~\ref{fig:deltaC}, where we have plotted maximum versus minimum count
rate. Constant sources would lie on the dotted line. 
The dashed line corresponds to a factor of 10 change.
\begin{figure}
\begin{center}
\resizebox{8cm}{!}{\includegraphics{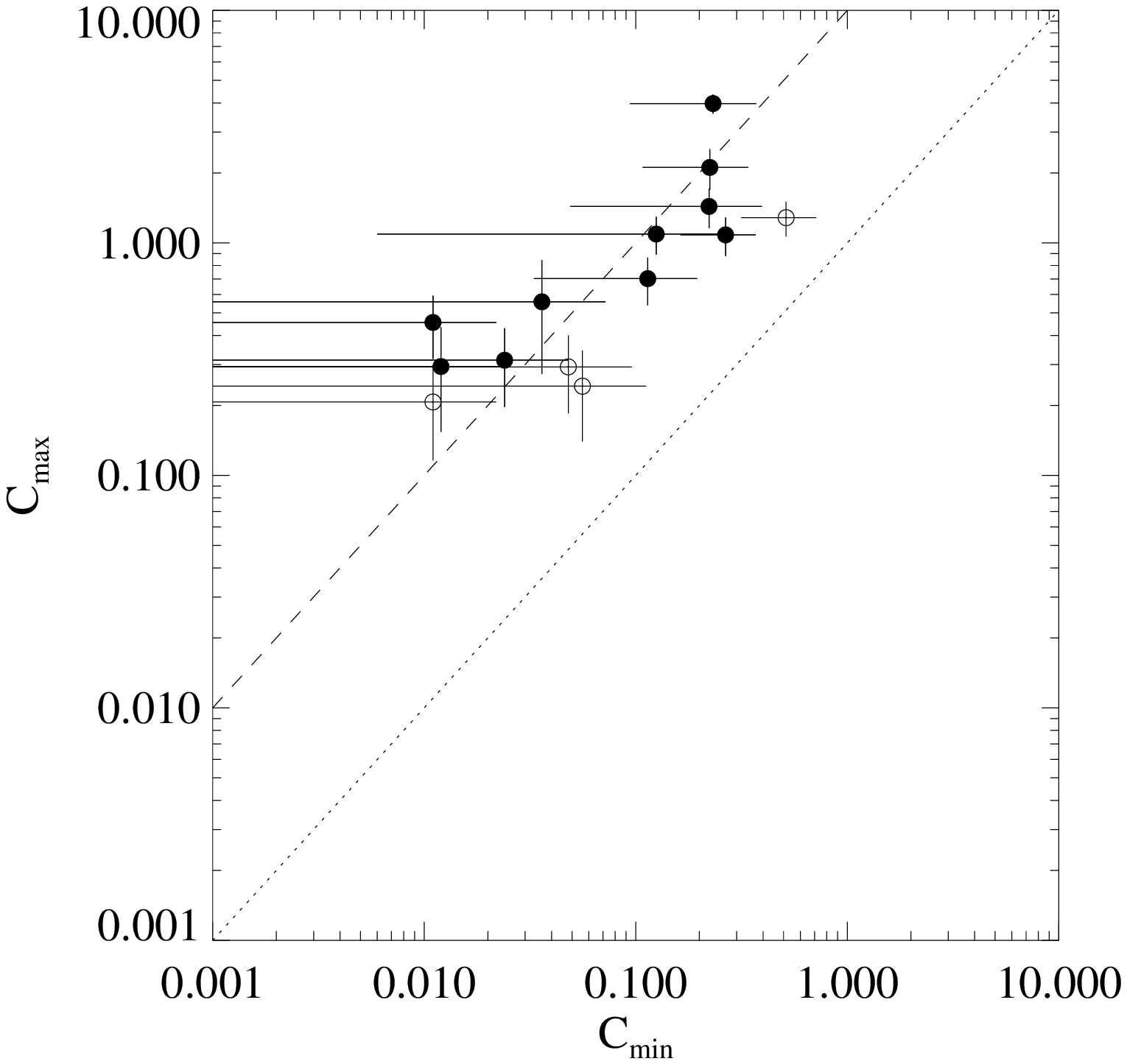}}
\caption{Count rate variations for individual stars during the RASS
observation. Displayed are the minimum (x-axis) and maximum (y-axis) count
rates observed in the RASS lightcurves. 
Variations which are significant at the $3\,\sigma$\,level are marked
by filled symbols.
The dotted line corresponds to a constant source, and the
dashed line represents a change by a factor of 10.}
\label{fig:deltaC}
\end{center}
\end{figure}

\begin{figure*}
\begin{center}
\resizebox{18cm}{!}{\includegraphics{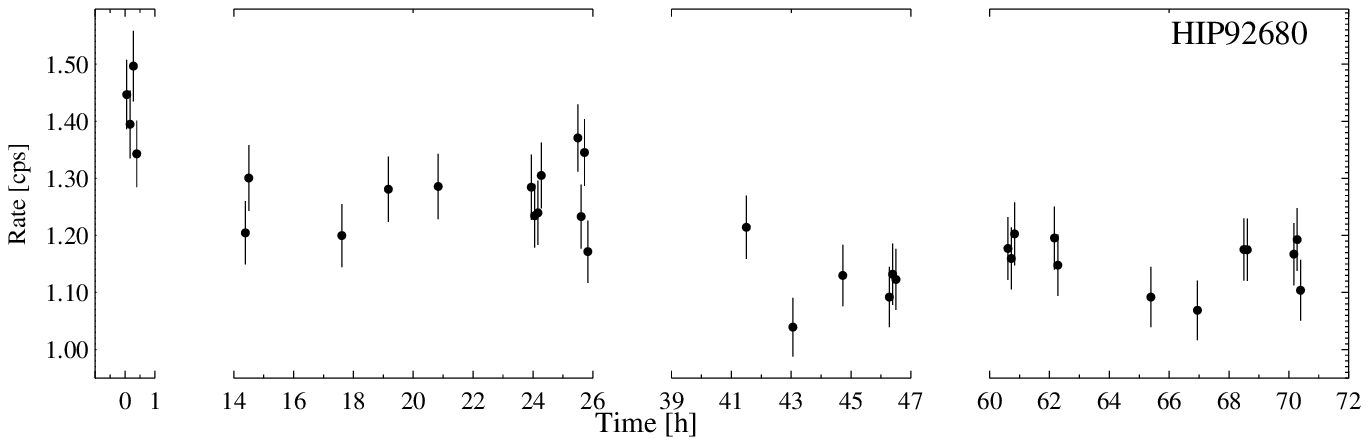}}
\caption{PSPC pointing lightcurve of HIP\,92680. Binsize 400\,s, 1\,$\sigma$ uncertainties.}
\label{fig:lcs_pspc}
\end{center}
\end{figure*}

Among the pointed PSPC observations, only the lightcurve of HIP\,92680 
shows significant variability. However, the other two observations
(ROR\,200099p and 200404p) 
are rather short, and therefore long-term variations on HIP\,100751 and 
HIP\,103438 (an improbable member) 
might have been missed. The lightcurve of HIP\,92680 is displayed in 
Fig.~\ref{fig:lcs_pspc}. Remarkably, this star was among the few sources
not found to be variable (at the 3\,$\sigma$ level) during the RASS (see 
Table~\ref{tab:deltaC}).

\section{Conclusions}\label{sect:conclusions}

We have searched the {\em ROSAT} All-Sky Survey for X-ray emission from
potential members of the recently identified Tucanae association. 59\%
of the stars labeled `probable' members by \citey{Zuckerman00.1} are
detected, but only 7\% of the `improbable' members. The RASS XLDF of the
probable members is very similar to the XLDF for the Taurus-Auriga
star forming region, and the young open cluster IC\,2602. The XLDF of the 
TW\,Hydrae association has the same $\langle \lg{L_{\rm x}} \rangle$, but shows
a somewhat smaller spread of luminosities. This is presumably due to 
the narrow range of spectral types among the TW\,Hydrae stars and/or
the assumed uniform distance of 55\,pc for all stars without measured parallax.
The similarity of the XLDF for the above mentioned regions indicates
that the X-ray emission does not change significantly with age from the 
early PMS stage (Taurus-Auriga) until the phase when the main-sequence (MS) 
is reached (IC\,2602).
However, the XLDF of the 100\,Myr old Pleiades cluster is characterized 
by significantly weaker X-ray emitters than all other samples and suggests
that once on the MS the stellar X-ray luminosity decreases.
From this comparison we infer an age between $10-30$\,Myr for the 
Tucanae association.

Most of the RASS detected Tucanae members have highly variable
lightcurves. The only star observed in a long PSPC pointed exposure
shows strong variations there, but was not significantly variable during
the RASS. This
strengthens the hypothesis that probably all Tucanae stars are 
variable given long enough observing time. The strong X-ray variability
observed can be considered another indicator for the youth of these systems.

The youth and close distance of the Tucanae stars makes them
good candidates for direct imaging of substellar companions, both brown
dwarfs and even giant planets, because substellar objects are hot and
bright when young (\cite{Burrows97.1}) and well separated when nearby.
I.e. they are detectable with the current technology
(e.g. \cite{Neuhaeuser00.1}). The Tucanae members are as well suited for
this purpose as the TW\,Hya members and the MBM\,12 T Tauri stars
(\cite{Hearty00.1}, \cite{Hearty00.2}).

\begin{acknowledgements}
This research has made use of the {\em ROSAT} All-Sky survey data which have 
been processed at MPE, and the Open Cluster Data Base 
at ftp://cfa-ftp.harvard.edu/pub/stauffer/clusters provided by J. Stauffer
and colleagues. 
The {\em ROSAT} project is supported by the
Max-Planck-Gesellschaft and Germany's federal government (BMBF/DLR).
\end{acknowledgements}

\end{document}